\documentclass[11pt]{article}
\usepackage{amsfonts}
\usepackage{amsfonts}
\usepackage{amsfonts,amsmath,amssymb,epsf,indentfirst}
\usepackage{graphicx}
\usepackage{pst-grad} 
\usepackage{verbatim}
\numberwithin{equation}{section}
\usepackage{hyperref}

\newcommand{\be}{\begin{equation}}
\newcommand{\ba}{\begin{eqnarray}}
\newcommand{\ea}{\end{eqnarray}}
\newcommand{\ee}{\end{equation}}

\newcommand{\we}{\wedge}

\newcommand{\f}{\frac}

\newcommand{\ti}{\tilde}

\newcommand{\no}{\nonumber \\}

\newcommand{\ep}{\epsilon}

\begin{document}

\begin{titlepage}
\thispagestyle{empty}

\begin{flushright}

\end{flushright}

\bigskip

\begin{center}
\noindent{\Large \textbf{$SL(2,\mathbb{Z})$ Action on AdS/BCFT and Hall Conductivities}}\\
\vspace{15mm} Mitsutoshi Fujita\footnote{email: mf29@uw.edu },  Matthias Kaminski\footnote{email: mski@uw.edu}, Andreas Karch\footnote{email: akarch@u.washington.edu }
\vspace{1cm}

{\it  Department of Physics, University of Washington, Seattle, WA 98195-1560, USA} \\

\vskip 3em

\begin{abstract}{We study the response of a conserved current to external electromagnetic fields in a holographic system with boundaries using the recently proposed AdS/BCFT~(boundary
conformal field theory) framework. This, in particular, allows us to extract the Hall current, the Hall conductivity, plus some potentially novel transport coefficients, and relations among them. We also analyze the action of SL(2,Z) duality in the gravity bulk, which acts non-trivially on the conductivity of the BCFT.
Finally we consider a type IIA string theory embedding of our setup.
}
\end{abstract}
\end{center}
\end{titlepage}

\tableofcontents

\section{Introduction}

The dynamics of a large class of strongly correlated toy models can be solved exactly using the tools of holography~\cite{Maldacena:1997re}. One of the interesting problems this idea can be applied to is charge transport in strongly coupled systems in response to external fields. In particular the Hall conductivity in such systems can be extracted either by studying linear response \cite{Hartnoll:2007ai} or by finding suitable stationary states in the presence of external fields \cite{Oba1}. For the former one is limited to small external electric fields but has access to DC as well as AC conductivities. In the latter case, one is restricted to the DC case but has access to non-linear phenomena in the presence of strong external fields. When simultaneously applicable the two methods agree.

Both these calculations have been performed for (field theory) bulk materials.\footnote{Note that
in the condensed matter community the bulk of a material is distinguished from its
edges. We refer to this bulk as the (field theory) bulk. In the context of AdS/CFT the
word bulk refers to the spacetime in which the gravity dual lives. We refer to this
as the (gravity) bulk.}
The physics of the quantum Hall effect however is deeply tied to the behavior of edge modes. The easiest way to incorporate edge modes in the context of AdS/CFT is by introducing defects. This is particularly easy in the so called probe limit, where in the field theory the number of degrees of freedom on the defect is much smaller than the number of degrees of freedom in the (field theory) bulk. So on the dual gravity side the defect does not backreact on the surrounding geometry \cite{Karch:2000gx}. Setups of this type have been used to construct the edge modes of integer and fractional quantum Hall states in \cite{Fujita:2009kw}\footnote{Holographic models of the quantum Hall effect have also been studied in~\cite{30,DKS,Belhaj:2010nn,Bergman:2010gm,Bayntun:2010nx,Bayntun:2011bm,Cadoni:2011kv}.}. While the case that the defect does backreact is difficult to study in the context of string theory, it has been understood in the context of Randall-Sundrum (RS) braneworlds \cite{Randall:1999vf}. As long as a defect can be holographically described by an infinitesimally thin brane with a Nambu-Goto like action,
a conformal field theory (CFT) with backreacting defects or even with boundaries can be easily realized \cite{Karch:2000gx,Karch:2000ct}. In the (gravity) bulk the jump in extrinsic curvature across the brane gets balanced by the tension of the brane. Only recently have full supergravity solutions been found that correspond to genuine string theory duals to defects with backreacting defects and boundaries \cite{D'Hoker:2007xy,D'Hoker:2007xz,Aharony:2011yc,Chiodaroli:2011fn}. The thin Randall-Sundrum brane of \cite{Karch:2000ct} can be seen as a ``bottom-up" model that captures the essential features of the thick branes found in these stringy constructions. This is analogous to how the hard-wall model of bottom-up QCD can capture the relevant features of a smooth supergravity solution dual to a confining gauge theory. In fact, the Randall-Sundrum perspective makes it clear that these two phenomena are closely related. From the (gravity) bulk point of view both are described by spacetime terminating at a brane. If the brane cuts off the (gravity) bulk spacetime in the radial direction, the dual CFT has a cutoff (a UV cutoff for positive tension branes, a hard-wall IR cutoff for negative tension branes). If the brane reaches the boundary, the corresponding cutoff is a spatial cutoff in the field theory: the field theory has a boundary.

In an elegant paper \cite{Takayanagi:2011zk}, Takayanagi recently built upon this earlier work and proposed a general construction for boundary CFTs (BCFTs) and their holographic duals; many details and examples have been worked out in \cite{Fujita:2011fp}. See also~\cite{Alishahiha:2011rg,Setare:2011ey,Setare:2012ks} for  two point functions in AdS/BCFT. These studies are once more in the context of thin branes. As we emphasized, these should generically be seen as a stand-in for a more complete description in terms of a supergravity dual\footnote{The one exception to this are orientifold planes, which even in the context of string theory are best described as a thin object with tension. The other important feature they have is that the tension of an orientifold can be negative, which will be crucial for some of our constructions.}. As long as the brane tension compensates for the extrinsic curvature of the spacetime, latter can consistently end on the brane. The variational principle for gravity on such a spacetime is still well defined, the brane effectively enforcing Neumann boundary conditions on gravitational fluctuations.

In this work we are including a (gravity) bulk Maxwell field in these holographic duals to 2+1 dimensional BCFTs. We show that for a subset of models one can easily construct a stationary solution describing the currents in this system in response to external fields. As we will demonstrate, the crucial feature (of the models in which this is possible) is that the 2+1 dimensional CFT effectively acquires a (position dependent) gap in response to the defect.
As a consequence the longitudinal conductivities vanish and we are only left with the topological Hall responses. In this sense our model is similar to the $(2+1)$-dimensional electron system
displaying a quantum Hall effect without a (large) magnetic field~\cite{Haldane:1988zza}.
Naively we would have expected that in our model transport is only driven by the edge modes which are not localized electrons but the flow of electrons in the direction of the cyclotron orbit on the defect.
However, if our transport truly was driven by the edge, we should see a current density that peaks close to the edge. Our current densities are independent of distance to the edge. They are
(field theory) bulk quantities. Nevertheless, the presence of
the edge modifies the Hall currents in an interesting way which we document.
The resulting Hall currents display either integer or fractional quantum Hall behavior depending on details of the model.

The purpose of this paper is to build a dual to quantum Hall
effects using AdS/BCFT, to analyze transport coefficients in this setup,
to reveal relations among them, and to study their transformation properties under
$SL(2,\mathbb{Z})$. Along the way we find potentially novel transport coefficients associated
with the magnetic field or with the time component of the conserved current.
{For a review of quantum Hall physics see~\cite{Zee:1996fe}.}

One motivation to study the action of $SL(2,\mathbb{Z})$ is the equivalence which is called a duality transformation~\cite{Girvin:1984zz,JKJain1990,JKJain1992} of Hall conductivities in the condensed matter literature.{\footnote{In the context of 
quantum Hall systems the modular group has been discussed
in~\cite{Fradkin:1996xb,Burgess:2000kj,Shapere:1988zv,Rey:1989jg,Lutken:1991jk,Kivelson:1992zz,Lutken:1992xj,Dolan:1998vr,Burgess:1999ug}.}} There, \emph{duality transformations} exists for the $(2+1)$-dimensional electron system in a magnetic field, with the filling fraction $\nu$, and if the electron-electron interaction is fixed:
\ba
&(i)\ \nu \leftrightarrow \nu +1, \label{DUA11} \\
&(ii)\ \nu \leftrightarrow 1-\nu \ \,\text{for}\, \ \nu <1, \label{DUA12} \\
&(iii)\ \dfrac{1}{\nu} \leftrightarrow \dfrac{1}{\nu}+2,\ \,\text{or}\, \ \nu  \leftrightarrow \dfrac{\nu}{2\nu +1}. \label{DUA13}
\ea
Note that the duality (iii) relates systems with an integer quantum Hall effect (IQHE) to 
ones with a fractional quantum Hall effect (FQHE). This is particularly interesting
because the IQHE can be understood in terms of free electrons, while the
FQHE requires interacting electron systems.
Corresponding relations for the electromagnetic response functions (such as the conductivities) are also available. $(i)$ is the Landau level addition transformation. $(ii)$ is the particle-hole transition implying that the state with the filling fraction $\nu$ of electrons is related with the same filling fraction of holes. $(iii)$ is the flux-attachment transformation. By using a mean field approximation, it is shown that attaching two flux quanta $h/e$ to each particle does not affect the physics. These actions on $\nu$ (or on the conductivities which can be given in terms of $\nu$) show a structure of
the form $\Gamma_0(2)$.
Here, $\Gamma_0(2)$ is a subgroup of $SL(2,\mathbb{Z})$ and is generated using two elements $ST^2S$ and $T$.\footnote{A different subgroup $\Gamma_{\theta}(2)$ of $SL(2,\mathbb{Z})$ is relevant to the quantum Hall effect with bosons as a fundamental particle.} Note that $\Gamma_0(2)$ does not include the single $S$-operation of $SL(2,\mathbb{Z})$ in its elements.

In the context of the AdS/CFT correspondence, it was found in~\cite{Bayntun:2010nx,Bayntun:2011bm} that  the $d=4$ axio-dilaton gravity with $SL(2,\mathbb{Z})$ symmetry
provides a holographic model of the mentioned duality transformation~\eqref{DUA11}--\eqref{DUA13} in the quantum Hall effect.
Their analysis is based on the $SL(2,\mathbb{Z})$ transformation acting on the usual AdS/CFT correspondence proposed in~\cite{Witten:2003ya}. It is known that the $S$ operation of $SL(2,\mathbb{Z})$ in the 3-dimensional CFT with $U(1)$ symmetry is considered as mirror symmetry and the $T$ operation is interpreted as shifting the Chern-Simons coupling. These actions of $SL(2,\mathbb{Z})$ in the 3-dimensional CFT are related with the $SL(2,\mathbb{Z})$ of 4-dimensional $U(1)$ gauge theory on $AdS_4$. Moreover, the application of the $SL(2,\mathbb{Z})$ symmetry for the condensed matter physics has been investigated including the models of the cyclotron resonance~\cite{Hartnoll:2007ip,Herzog:2009xv,Goldstein:2010aw}, the particle-vortex duality called mirror symmetry~\cite{Herzog:2007ij}. It was found in~\cite{Hartnoll:2007ip}  that the conductivity $\sigma =\sigma_{xy}+i\sigma_{xx}$ computed via holography transforms under
$SL(2,\mathbb{Z})$ as
\ba
\sigma =\dfrac{\hat a\sigma +\hat b}{\hat c\sigma +\hat d}, \label{DUA14}
\ea
where $\hat a,\hat b,\hat c,\hat d\in \mathbb{Z}$ and $\hat a\hat d-\hat b\hat c=1$. In our present paper, on the other hand, we impose
boundary conditions on the $U(1)$ gauge fields at the additional boundaries in the AdS spacetime.
Thus, we expect that the $SL(2,\mathbb{Z})$ transformation of the conductivity \eqref{DUA14} is changed in the presence of the additional boundaries.

This paper is organized as follows: In section \ref{sec:FQHE}, we briefly review the
$d=4$ gravity dual to the half plane. In this background, we introduce the $U(1)$ effective action,
derive the conductivity in the presence of the additional boundary,
and consider boundary conditions of both Dirichlet and Neumann type at the additional boundary.
In section \ref{sec:SL(2,Z)}, we extract the Hall current and find (partly novel) transport coefficients.
We further apply the $SL(2,\mathbb{Z})$ transformation in the (gravity) bulk spacetime and derive the  transformation properties of the conductivity on the BCFT side under this $SL(2,\mathbb{Z})$.
In section \ref{sec:stringy}, we give the type IIA string theory embedding of the $U(1)$ effective action,
concluding with a brief discussion in section \ref{sec:discussion}.

\section{FQHE from the gravity dual to BCFT}\label{sec:FQHE}

In this section we construct a holographic setup which realizes a
BCFT with a conserved current. We extract the components of this current
and derive transport coefficients, most prominently the Hall conductivities
from them. Finally, an Onsager relation for the Hall conductivities and
the ensuing restrictions on our model are discussed.

\subsection{Holographic setup}

According to~\cite{Takayanagi:2011zk,Fujita:2011fp}, we consider the $AdS_4$ gravity dual of the BCFT on the half plane $y>0$, where the dual BCFT  lives on 3-dimensional space parametrized by $t,x,y$ ($y>0$).
We start with the 4-dimensional Einstein-Hilbert action with the boundary term
\ba
I=\dfrac{1}{16\pi G_N}\int d^4x \sqrt{-g}(R-\Lambda)+\dfrac{1}{8\pi G_N}\int_{Q}d^3x\sqrt{-\gamma}(K-T), \label{ACT11}
\ea
where $\Lambda =-6/R^2$, $K$ is the extrinsic curvature, and $T$ is the tension of the boundary $Q$. We denote the induced metric on $Q$ as $\gamma_{\mu\nu}$.
We consider the $AdS_4$ spacetime with $3$-dimensional boundary $Q$, where isometry $SO(2,3)$ is broken into $SO(2,2)$ in the presence of $Q$.
Computing the variation of \eqref{ACT11} at the boundary $Q$,  we find that terms including the derivative of the $\delta\gamma_{\mu\nu}$ cancel in the presence of the second term. We then find the following boundary condition:\footnote{We ignored the Gibbons-Hawking term on the $AdS_4$ boundary. If we compute the energy momentum tensor of the dual BCFTs,
    we should add this term and impose the Dirichlet boundary condition on $\delta g_{\mu\nu}$.}
\ba
K_{\mu\nu}=(K-T)\gamma_{\mu\nu}.
\ea

\begin{figure}[t]
\begin{center}
\vskip -1.5cm
\includegraphics[width=14cm]{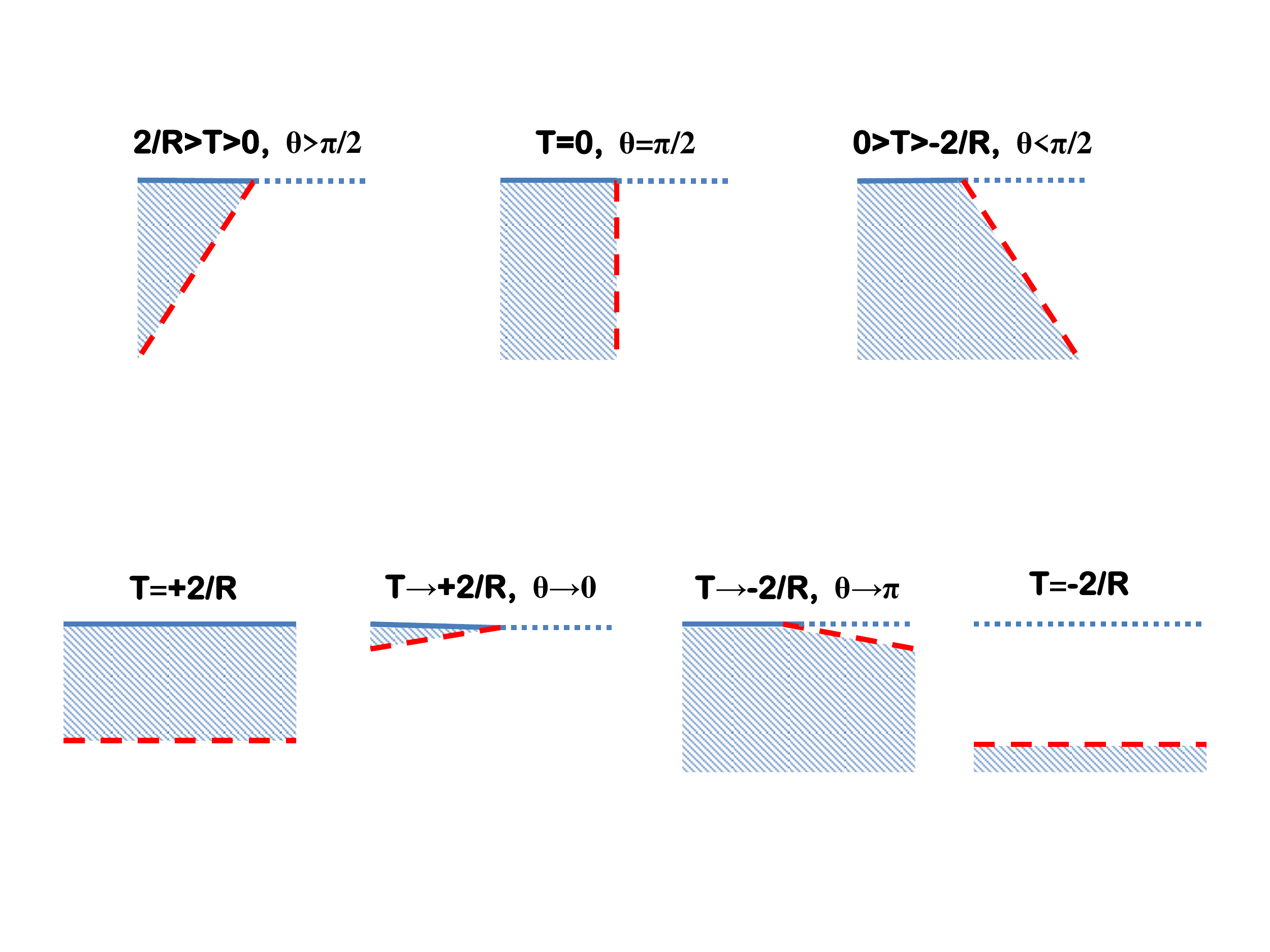}
\vskip -1.5cm
\caption{\label{fig:adsbcft} Embedding of the defect brane corresponding to various values of the tension $T$ and
hence to various angles $\theta$. The solid (blue) line is the part of AdS boundary on which the BCFT lives. The dashed (red) line the brane $Q$ at which spacetime terminates. The filled part of spacetime is excised; the unfilled part is physical. The dotted (blue) line is the part of the AdS boundary that is cut out.}
\end{center}
\end{figure}

The $AdS_4$ metric  is given by
 \ba
ds^2=R^2\dfrac{-dt^2+dx^2+dy^2+dz^2}{z^2}.
\ea
Here, the field theory spacetime is restricted to the half plane $y>0$ at the $AdS$ boundary in the presence of $Q$. The (gravity) bulk spacetime dual to the half-plane is restricted to the $AdS_3$ region inside $Q$,
\ba
Q: y=\cot \theta z, \quad \quad 0 < \theta < \pi,
\ea
where the vector normal to $Q$, pointing to the outside of the gravity region, is given by $n_{\mu}=(0,0,-\sin\theta,\cos\theta)$ and the vector parallel to $Q$ is given by $l_{\mu}=(0,0,\cos\theta,\sin\theta)$. We have sketched the corresponding spacetimes in figure \ref{fig:adsbcft}.

The extrinsic curvature $K_{\mu\nu}$ and the tension $T$ for the above solution have the following form:
\ba
K_{\mu\nu}=-\dfrac{\cos\theta}{R}\gamma_{\mu\nu}, \quad T=-\dfrac{2\cos\theta}{R}.
\ea
The second equation means that the tension is restricted to the region
$-2/R \le T \le 2/R$. For tensions taking the values at the two ends of this interval the brane is described by the equation $z=0$, which means that there is defect in the theory any more.
In this case the condition on the extrinsic curvature is in fact obeyed for any equation of the form $z=z_*$. For example for $T=-2/R$ the vector normal to this boundary pointing to outside of the gravity region is given by $N_{\mu}=(0,0,0,R/z)$ satisfying $N_{\mu}N^{\mu}=1$.
Using the induced metric $\gamma_{\mu\nu}=\text{diag}(-R^2/z^2,R^2/z^2,R^2/z^2,0)$,
the extrinsic curvature $K_{\mu\nu}$ and the tension $T$ for the above solution have the following form:
\ba
K_{\mu\nu}=-\dfrac{1}{R}\gamma_{\mu\nu}, \quad T=-\dfrac{2}{R}.
\ea
For the negative tension $T=-2/R$ brane this is just the standard hard wall~\cite{Erlich:2005qh}, or the infrared RS brane, corresponding to an IR mass scale in the theory. The $z>z_*$ region of spacetime is removed. In the case of the positive tension $T=2/R$ brane we have a UV RS brane corresponding to a CFT with a UV cutoff; the AdS boundary is completely excised in that case\footnote{When the magnitude of tension is larger than $2/R$ one finds "de-Sitter RS" branes; their embedding is given by an equation of the form $t \sim z$ instead of $y \sim z$. In terms of the holographic duals they correspond to field theories with a boundary in the time direction, as appropriate for a quantum quench. It would be interesting to explore this further.}. Both of these special $T=2/R$ cases are also displayed in figure \ref{fig:adsbcft}. Note that if we consider the BCFT defined in the half plane $y<0$ instead of $y>0$, we have different signs of the normal $n_{\mu}$ and the tension as expected.

Defining the coupling constants $c_1=1/2g^2$ and   $c_2=\Theta/8\pi^2$, we introduce a
(gravity) bulk $U(1)$ gauge field with the following action to describe the Hall effect:
 \ba \int c_1*F\wedge F+c_2F\wedge F-f(\phi )\int_{Q} A\wedge F+V(\phi), \label{ACT130} \ea
 where $Q$ is the region defined above, $*$ is the epsilon symbol in $AdS_4$, and $\eta$ is a constant. In \eqref{ACT130}, the vev of the scalar field $\phi$ fixed by the potential $V(\phi)$ 
determines the Chern-Simons level $f(\phi)$ which in particular cases (see discussion
in section~\ref{sec:discussion}) can be interpreted as the strength of the magnetic field perpendicular to the $d=2$ space directions. In a description of Hall experiments usually the electric density is fixed and the magnetic flux is varied.
Note that the integration of the topological term gives a Chern-Simons term with negative sign at the boundary $Q$, and also at the AdS boundary.

Recall that the Chern-Simons theory at $Q$ is not gauge invariant in the presence of the boundary $\partial Q$.
To make the action gauge invariant, we include a boundary term at $\partial Q$~\cite{Elitzur:1989nr,Jensen:2010em,Fujita:2009kw} to \eqref{ACT130} as follows:
  \ba \int c_1*F\wedge F+c_2F\wedge F-f(\phi )\Big(\int_{Q} A\wedge F+\eta \int_{\partial Q}d^2xA_tA_x\Big)+V(\phi). \label{ACT13} \ea
The gauge transformation $\delta A=d\chi$ of \eqref{ACT13}  is given by
 \ba
& -f(\phi)\delta \Big(\int_{Q} A\wedge F +\eta\int_{\partial Q}d^2xA_tA_x\Big) \nonumber \\
& =-f(\phi)\int_{\partial Q} d^2x\Big[(-1+\eta) \delta A_tA_x +(1+\eta)\delta A_x A_t\Big] \nonumber \\
&=f(\phi)\int_{\partial Q} d^2x \Big[(-1+\eta) \chi \partial_t A_x +(1+\eta)\chi \partial_x A_t\Big],
 \ea
where setting $\eta=-1$\footnote{Different choice $\eta= 1$ realizes the non-zero electric field with the condition $\partial_x A_t=0$.}, the condition $\partial_t A_x=0$ makes the action gauge invariant and can still realize the non-zero electric field $F_{tx}\neq 0$. Moreover, we should impose the boundary condition for the variation $\delta A_t|_{\partial Q}=0$.

\subsection{An example for Hall physics} \label{sec:example}

A typical setup for Hall experiments in 2+1 dimensions involves a current $J^x$ perpendicular to
an external electric field $E_y$ in presence of a magnetic field $B$\footnote{Note that the latter
is a pseudoscalar in 2+1 dimensions (not a pseudovector as in 3+1 dimensions).}.
We are now looking for a particular ansatz for our
gauge fields $A_{\mu}$, which realizes this field configuration in our model. The gauge
fields also have to obey the equation of motion obtained by varying the action \eqref{ACT13} with respect to $A$, $d(*F)=0$. One adequate ansatz for the vector fields $A$ is
 \ba \label{eq:ansatz}
 A_t= E_y y+bz,\quad A_x= -B y+cz,\quad A_y=dz,\quad A_z=0, \label{ANS28}
 \ea
where the parameters $b,c,d$ are related to the current density $\langle J^{\mu}\rangle$ via the GKPW \cite{Gubser:1998bc,Witten:1998qj} ansatz. Note that the above solution satisfies the condition $\delta A_t|_{\partial Q}=0$.

We should pause for a moment and consider the case without a boundary, that is when the spacetime is pure AdS. In this case we are describing a CFT living on flat 2+1 dimensional space. In the solution (\ref{eq:ansatz}) we have the background electric and magnetic fields as well as the charge density and the currents as free parameters. Clearly they can not all be independent. We know for example that in the presence of finite density a finite electric field will typically lead to a run away current. No static solution should exist whatsoever. The problem with our ansatz is that we have yet to implement an IR boundary condition, that is a condition at $z=\infty$. Calculating the Lagrangian density on our ansatz we find that it is a constant; the AdS$_4$ gauge field action is just the same as the flat space gauge field action and clearly constant fields have a finite action proportional to $\vec{E}^2 - \vec{B}^2$. But the action itself will be infinite for such a field configuration, as we are integrating a finite density over an infinite volume. Note that the infinity from integrating over the field theory dimensions is physical and can easily be regulated by putting the system at a finite volume. As expected, one finds an energy extensive in volume. However the infinity from integrating over the infinite range of $z$ should give us a pause. This corresponds to an infinite energy density at every point in field theory space. This is a singularity of the above solution and tells us it should be discarded as long as $z=\infty$ is part of the spacetime. By choosing $c^2+d^2+B^2=E^2+b^2$ one can make the Lagrangian density vanish identically, but presumably this is still not a well behaved gauge field configuration. For example one would expect $A_t$ to vanish at the horizon as is always imposed in the study of holographic finite density systems.

In order to determine what the right boundary condition at the horizon $z=\infty$ is, one needs to have the full non-linear action of the (gravity) bulk gauge field. For example in the case that the
(gravity) bulk gauge field is given by a Born-Infeld action as appropriate for a probe D-brane, requiring that the action remain real completely fixes the currents in terms of the background fields \cite{Kar1,Oba1,Karch:2010kt}. As we will see below the details of this boundary condition are not important for us; as soon as the region $z=\infty$ is part of our (gravity) bulk spacetime any additional boundary condition will invalidate the ansatz (\ref{eq:ansatz}).

\paragraph{Regularization of the action}
As seen in the usual AdS/CFT case, we can achieve the regularization of the on-shell action (using the equation of motion). The relevant part of the action is given by
\ba
&I_{on-shell}=\int c_1*F\wedge F+c_2F\wedge F -f(\phi)\Big(\int_{Q}A\wedge F -\int_{\partial Q_1} dx^2A_tA_x \Big) \nonumber \\
&=\int_{z\to 0} A\wedge (c_1*F+c_2F) -\int_{Q}A\wedge (c_1*F+(f(\phi)+c_2)F) \nonumber \\
&=\int_{z\to 0} A\wedge (c_1*F+c_2F), \label{REG137}
\ea
where in the second line, we performed the integration by parts and used the EOM $d(*F)=0$. Note that the boundary term at $\partial Q_1$ in the first line becomes zero using the solutions \eqref{ANS28}.
In the third line, we required the contribution of the boundary $Q$ to vanish. This may be achieved
by a Dirichlet or alternatively a Neumann boundary condition as explained in the next paragraph. Note that the above on-shell action can be obtained for any value of  $\theta$.
According to~\cite{Henningson:1998gx,Henningson:1998ey,Kar1}, the kinetic term in the third line of \eqref{REG137} can be renormalizable holographically.

\paragraph{Boundary conditions}
When we include the defect we also have to impose the following boundary condition at $Q$\footnote{Using the differential form and restricting the variation of $A$ to the boundary, the boundary condition is given by $c_1*F+(c_2+f(\phi))F|_{Q}=0$.}:
\ba
-c_1\sqrt{-g}n_{\nu}F^{\nu\mu}+\dfrac{(f(\phi)+c_2)}{2}\epsilon^{\mu\nu\rho}F_{\nu\rho}|_{Q}=0, \label{BOU18}
\ea
where $n_{\mu}$ is the normal of $Q$ and $\epsilon^{\mu\nu\rho}$ is the flat epsilon symbol in 3 dimensions. The boundary condition \eqref{BOU18} is interpreted as the Neumann boundary condition, and can be rewritten as
\ba
&0=c_1(B\sin\theta +c\cos\theta )-(f(\phi)+c_2)(E_y \cos\theta +b\sin\theta ), \label{BOU16} \\
&0=c_1(b\cos\theta -E_y \sin\theta )+(f(\phi)+c_2)(B\cos\theta -c\sin\theta ),\quad c_1d=0. \nonumber
\ea

The solutions in terms of $b,\,c$ and $d$ are given by
\ba
&b=-{\dfrac {(f \left( \phi \right)+c_2)B{ c_1}- (\left( f \left( \phi \right) +c_2 \right) ^{2}+c_1^2)\cos
 \theta \sin \theta  E_y}{{{c_1}}^{2}\cos^2
 \theta- \left( f \left( \phi \right)+c_2
 \right) ^{2} \sin^2\theta }}, \\
&c=-{\dfrac {({{ c_1}}^{2}+(\left( f \left( \phi \right)+c_2
 \right) ^{2})B\cos \theta \sin \theta
-(f \left( \phi \right)+c_2) {
\it c_1}\,E_y}{{{c_1}}^{2}\cos^2
 \theta- \left( f \left( \phi \right)+c_2
 \right) ^{2} \sin^2\theta }}, \quad &d=0.
\ea

If we assume that $f(\phi)$ is constant, then we can use the above condition for any $y$. Note that there were 5 undetermined constants $E_y,B,b,c,d$ and 3 independent equations \eqref{BOU16}. If we regard the electric
and magnetic fields at inputs which are specified at the AdS boundary $z \to 0$, imposing a Neumann boundary condition similar to \eqref{BOU18} completely determines the gauge field solution $A_{\mu}$.
This situation is analogous to the case of 1-point functions of the stress tensor given in~\cite{Fujita:2011fp}. The current density is derived using the GKPW relation as follows:
\ba
&J^t=\dfrac{\delta S}{\delta A_t}\Big|_{bdy}=2\epsilon_{(3)}(c_1*F+c_2F)|_{bdy}=2c_1b+2c_2B, \label{GKP11} \\
& J^x=\dfrac{\delta S}{\delta A_x}\Big|_{bdy}=2\epsilon_{(3)}(c_1*F+c_2F)|_{bdy}=-2c_1c+2c_2E_y, \nonumber \\
& J^y=\dfrac{\delta S}{\delta A_y}\Big|_{bdy}=0, \nonumber
\ea
where the functional derivative is defined in the AdS boundary $z\to 0$ and $\epsilon_{(3)}$ represents $\epsilon^{\mu\nu\rho}$. Here, the convention of the differential form is used at the AdS boundary.

So the form of the solution is completely fixed by the boundary conditions at $Q$. But as long as the region $z=\infty$ is part of the (gravity) bulk spacetime we have one more boundary condition to impose. As the solution is already fixed, any additional boundary condition will not be met. Instead the ansatz eq.(\ref{eq:ansatz}) has to be abandoned.
Note that eq.(\ref{eq:ansatz}) can only capture {\it static} configurations for which the current density is {\it uniform} in space. Being forced to abandon eq.(\ref{eq:ansatz}) simply means no static, spatially homogenous solution exists in that case. In particular, as soon as we have a finite density in the system we would not expect the system to be static when external electric fields are turned on. What is required to obtain quantum Hall physics is a mass gap, and so far we are studying a CFT with boundary.

The only reason not to completely discard the ansatz eq.(\ref{eq:ansatz}) is that for the range of tensions $-2/R\leq T<0$ ($0\leq \theta<\pi/2$) the physical region of spacetime does {\it not} include $z=\infty$ ($z=\infty$ is only reached at $y=\infty$, so any boundary condition imposed there could be viewed as a boundary condition at infinity in the field theory as well which is not a problem). The backreaction of the defect was sufficient to gap the bulk field theory. From the profile of the brane solution we can see that the field theory cutoff falls off as one over distance to the boundary. This particular spatially varying gap apparently is just what is needed in order to support a spatially constant current density. In the extreme case of $T=-2/R$ we describe a gapped 2+1 dimensional system, the natural arena to expect quantum Hall physics. We will hence restrict our analysis to this interesting region of negative tensions $-2/R\leq T<0$.

A particularly interesting case to consider is $\theta=\pi/2$. Strictly speaking, for $\theta=\pi/2$ the $z=\infty$ region is not excised and we should think of $\theta=\pi/2$ only as a limiting case that can be approached from below. In this limit one has
\ba
&J^t=\dfrac{2c_1^2B}{f(\phi)+c_2}+2c_2B, \nonumber \\
&J^x=\dfrac{2c_1^2E_y}{f(\phi)+c_2}+2c_2E_y. \label{CUR112}
\ea

If we choose $c_1=1/4\pi$, $f(\phi)=k/4\pi$ and $c_2=l/4\pi$, this system seems to describe the FQHE of the filling fraction $\nu = \frac{1}{k+l}+l$ in units of $e=1$ and $\hbar =1$:
\ba
&J^t=\Big(\dfrac{1}{k+l}+l\Big)\dfrac{B}{2\pi}, \\
&J^x=\Big(\dfrac{1}{k+l}+l\Big)\dfrac{E_y}{2\pi} .
\ea
Under this choice, using Neumann boundary conditions, and $\theta=\pi/2$,
the Hall conductivity $J^x/E_y$ is given by
\begin{equation}\label{eq:sigmaxyThetaPi/2}
\sigma_{xy} = \Big(\dfrac{1}{k+l}+l\Big)\dfrac{1}{2\pi} \, .
\end{equation}
In summary, we have successfully constructed a system which shows the known fractional
quantization of the Hall conductivity.
Note that the case of filling fraction $\nu=5/2$ is of particular interest to the
condensed matter community.\footnote{See discussion in section \ref{sec:discussion}.}
For us $\nu=5/2$ is special in the sense that it is realized here
at vanishing Chern-Simons coupling $k=4\pi f(\phi)=0$ on the boundary $Q$
and with parameter $l=2$.

A second interesting special case for the boundary position is $\theta=0$.
Here the current \eqref{CUR112} becomes
\ba
&J^t=-{2(f(\phi)+c_2)B}+2c_2B=-k\dfrac{B}{2\pi}, \label{CUR116} \\
&J^x=-{2(f(\phi)+c_2)E_y}+2c_2E_y=-k\dfrac{E_y}{2\pi}. \nonumber
\ea
Here we are describing a gapped bulk field theory with a position independent gap and see exactly the standard Hall physics associated with a Chern-Simons term on the boundary\footnote{This current can also be obtained by noting that the boundary condition \eqref{BOU18} has the same form as \eqref{GKP11} except for the term with the coefficient $f(\phi)$.
After imposing \eqref{BOU18}, thus, we find the conductivity proportional to $f(\phi)$ as seen in \eqref{CUR116}.}. So with Neumann boundary conditions and for $\theta=0$, the Hall conductivity $J^x/E_y$ is
\begin{equation}\label{eq:sigmaxyTheta0}
\sigma_{xy} = -k\frac{1}{2\pi}\, . 
\end{equation}
Note that $\sigma_{xy}$ can take negative values for the Hall physics. This is different
from the longitudinal conductivity, which assumes only positive values.

It would be interesting to apply the analysis of this section to the gravity dual to the theory
with an interval $0<y<L$. This may be constructed by using two thin branes $Q_1$ and $Q_2$.
Naively two parallel branes with $\theta_1=\theta_2$ are a good solution. However one should note that in this case $z\to\infty$ is still part of the gravity dual. For Euclidean $AdS$ this would correspond to an extra boundary point and so the CFT seems to live on an interval plus an extra  lattice point. Such a point exists in the (gravity) bulk if two thin branes do not intersect. Presumably the correct solution one needs is one where the two defects smoothly connect. Solutions of this type have been constructed in \cite{Fujita:2011fp}. In this case the (gravity) bulk spacetime has to be modified in order to accommodate the presence of the defect.

\subsection{Dirichlet boundary condition on $Q$} \label{sec:DirichletBC}
Instead of the Neumann boundary condition \eqref{BOU18} at $Q$, we can consider the Dirichlet boundary condition at $Q$.
The Dirichlet boundary condition implies that the longitudinal components of the field strength for $Q$ become zero as follows:
\ba
n_{\mu}\tilde{F}^{\mu\nu}|_{Q}=0, \label{DIR31}
\ea
where $n_{\mu}=(0,0,-\sin\theta,\cos\theta)$.
With the ansatz \eqref{eq:ansatz} the above boundary condition is rewritten as
\ba
\cos\theta B-c\sin \theta=0,\quad E_y\cos\theta+b\sin\theta=0,\quad  F_{tx}=0.
\ea
Moreover, the regularization of the on-shell action gives another restriction. The condition of regularity can be written as
\ba
A_{\mu}|_{Q}=0, \label{BOU33}
\ea
where $\mu$ takes the values of coordinates on $Q$.
The above condition gives $d=0$. The boundary condition \eqref{BOU33} can be interpreted as the IR boundary condition for $T<0$ since the gravity theory ends at the boundary $Q_1$ with the maximal value of $z$.

The current at the boundary is given by
\ba
&J^t=-2c_1\cot\theta E_y+2c_2B,\quad J^x=-2c_1B\cot\theta +2c_2E_y, \nonumber \\
&J^y=0.
\ea
So with Dirichlet boundary conditions and for any $\theta$ the Hall conductivity $J^x/E_y$ becomes
\ba
\sigma_{xy}=-2c_1\dfrac{B}{E_y}\cot\theta +2c_2. \label{DIR34}
\ea

In order to understand how special the previous constructions with particular choices of
boundary conditions and special angles $\theta=0,\, \pi/2$ are, we will consider a more general case in
the coming section.

\subsection{Generalization and transport coefficients} \label{sec:conductivities}
In the previous subsections we have chosen a particular field configuration solving the
Maxwell equations. Here we generalize this configuration to include all possible constant
field strengths. This will enable us to investigate all the transport coefficients which
parametrize our system's response to external electric and magnetic fields.
The most general constant field strength is
\begin{eqnarray}
F_{zt}=b,\, && F_{xt}=E_x,\,  \nonumber \\
F_{zx}=c,\, && F_{yt}=E_y, \\ \nonumber
F_{zy}=d,\, && F_{xy} = -F_{yx} = B \, .
\end{eqnarray}
 One ansatz realizing this on the level of vector fields $A$ is
 \ba \label{eq:generalAnsatz}
 A_t=E_x x + E_y y+bz,\quad A_x= -B y+cz,\quad A_y=dz,\quad A_z=0,
 \ea
where the parameters $b,c,d$ are again related to the current density $\langle J^{\mu}\rangle$,
and this ansatz again satisfies the condition $\delta A_t|_{\partial Q}=0$.
\footnote{Note that it is not possible to add time-dependent terms linear in $t$
to the ansatz \eqref{eq:generalAnsatz} if we choose the boundary term with $\eta =1$ in \eqref{ACT13}. Such an ansatz would not respect our gauge choice $\partial_t A_x = 0$.
However, a time-dependent solution may be found in analogy to~\cite{Karch:2010kt}.
}

Using the general ansatz \eqref{eq:generalAnsatz}, we proceed completely analogously
to the previous subsection. We again impose the Neumann boundary condition \eqref{BOU18}
in order to obtain the more general $b,\, c$ and $d$.
Correspondingly, the current components are now given by
\begin{eqnarray}\label{eq:generalCurrent}
J^t & = &
2 {B} {c_2}+ 2\frac{{c_1} \left(-2 {B} {c_1} ({c_2}+f(\phi))+{E_y} \left({c_1}^2+({c_2}+f(\phi))^2\right) {\sin 2\theta}\right)}{2 {c_1}^2 {\cos\theta}^2-2 ({c_2}+f(\phi))^2 {\sin\theta}^2}
\, , \nonumber \\
J^x & = &
2 {c_2} {E_y}+\frac{{c_1} \left(-2 {c_1} {E_y} ({c_2}+f(\phi))+{B} \left({c_1}^2+({c_2}+f(\phi))^2\right) {\sin 2\theta}\right)}{{c_1}^2 {\cos\theta}^2-({c_2}+f(\phi))^2 {\sin\theta}^2}
\, , \\ \nonumber
J^y & = &
2 E_x f(\phi)
\, .
\end{eqnarray}

From the $x$-component of the corresponding current, we define the following transport coefficients
\ba\label{eq:JxCoefficients}
\sigma_{xx} &=& \frac{J^x}{E_{x}} = 0\, ,\nonumber\\
\sigma_{xy} &=& \frac{J^x}{E_{y}} =
2 {c_2} -\frac{2 {c_1}^2 ({c_2}+f(\phi))}{{c_1}^2 {{\cos}\theta}^2-({c_2}+f(\phi))^2
{\sin\theta}^2}  \, ,\\ \nonumber
\kappa_{x} &=& \frac{J^x}{B} = \frac{c_1({c_1}^2 + ({c_2} + f(\phi))^2) \sin2\theta}{{c_1}^2{\cos\theta}^2- ({c_2}+f(\phi))^2 {\sin\theta}^2} \, .
\ea
From the $y$-component of the current we get
\ba\label{eq:JyCoefficients}
\sigma_{yx} = \frac{J^y}{E_{x}} = 2 f(\phi) \, ,\qquad
\sigma_{yy} = \frac{J^y}{E_{y}} = 0\, ,\qquad
\kappa_{y} = \frac{J^y}{B} = 0 \, .
\ea
There are three more non-trivial coefficients that we can compute from the current
in $t$-direction
\begin{equation} \label{eq:JtCoefficients}
\gamma_x=\frac{J^t}{E_{x}} = 0  \, ,\qquad
\gamma_y=\frac{J^t}{E_{y}} = \kappa_x
 \, ,\qquad
\kappa_{t} = \frac{J^t}{B} = \sigma_{xy} \, .
\end{equation}
To our knowledge some of the relations \eqref{eq:JtCoefficients} have not been noted
before. In fact the coefficients $\kappa_x,\,\kappa_y$ and $\gamma_x,\, \gamma_y$
do not seem to have been investigated from this point of view yet. We discuss them
together with $\kappa_t$ below.

\paragraph{Onsager relations and consistency of our ansatz}
Let us explore if there are more relations among the transport coefficients found in the previous
paragraph.
For rotationally invariant theories we would immediately know $\sigma_{xy}=\sigma_{yx}$.
However, our boundary
at $Q$ breaks rotational symmetry in the $x,\,y$-plane, and so does our
electric background field unless we choose a symmetric ansatz such as
$A_t = E (x + y) + b z$.
But we know that $\sigma_{xy}$ and $\sigma_{yx}$
are related to each other by an Onsager relation for the two-point functions of currents \cite{PhysRev.37.405,PhysRev.38.2265}.
Namely, these two-point functions are related to transport coefficients by Kubo formulae~\cite{JPSJ.12.570},
in particular for the off-diagonal (Hall) conductivities we have schematically $\langle J^x J^y\rangle \sim \sigma_{xy}$ and $\langle J^y J^x\rangle \sim \sigma_{yx}$.

Now Onsager relations are based on the transformation properties of two-point functions
under time-reversal (and do not directly rely on rotational invariance). Although our theory
breaks time-reversal invariance, we can still utilize Onsager relations. The general
relation for two-point functions in Fourier space is $\langle \mathcal{O}_1\mathcal{O}_2\rangle (\omega, k, \mathfrak{p}) = n_1 n_2 \langle \mathcal{O}_2\mathcal{O}_1\rangle (\omega, -k, -\mathfrak{p})$, where $n_i$ is the eigenvalue of $\mathcal{O}_i$ under time-reversal. The array
$\mathfrak{p}$ represents all the parameters of our theory which break
time-reversal invariance.
In our case this amounts to the condition
$\langle J^x, J^y\rangle (\omega, k, \mathfrak{p}) = (-1) (-1)\langle J^y, J^x\rangle (\omega, -k, -\mathfrak{p})$, or in other words
\begin{equation}\label{eq:onsager}
\sigma_{xy} (\omega, k, \mathfrak{p}) = \sigma_{yx} (\omega, -k,
-\mathfrak{p})\, .
\end{equation}

Let us apply this Onsager condition to the off-diagonal conductivities found in
\eqref{eq:JxCoefficients} and \eqref{eq:JyCoefficients} under the assumption
that $\mathfrak{p} = \{c_2,\,f(\phi)\}$ are the only parameters which break
time-reversal invariance in our action \eqref{ACT13}. This would imply
\begin{eqnarray}\label{eq:explicitOnsager}
2 {c_2} -\frac{2 {c_1}^2 ({c_2}+f(\phi))}{{c_1}^2 {{\cos}\theta}^2-({c_2}+f(\phi))^2
{\sin\theta}^2} &=& -2 f(\phi) \, ,
\end{eqnarray}
which is in general {\it not} satisfied.

Consequently, our model at generic couplings can not describe a time-reversal invariant system. If there is no spontaneous time-reversal breaking in our system, our identification of time-reversal breaking parameters
$\mathfrak{p} = \{c_2,\,f(\phi)\}$ is complete. In this case we have to require that
\eqref{eq:onsager} be satisfied. This dictates a specific relation between the
parameters $c_1,\, c_2,\, f(\phi)$ and the angle $\theta$. We discuss possible
cases below.

For generic values of the couplings the only consistent interpretation of our results is that
the ground state of our
theory spontaneously breaks time-reversal. In fact the position-dependent mass
gap of our theory may be the result of such a spontaneous symmetry breaking.
In this case the Onsager relation as written in \eqref{eq:onsager} does not apply,
because we can not account for all sources of time-reversal breaking by flipping
the corresponding parameters in the microscopic Lagrangian.
We would have to change the boundary conditions on our fields
(in an unknown way) in order
to account for a symmetry-breaking mass gap.

Let us see what set of couplings could correspond to a time-reversal invariant theory
(modulo inversion of the time-reversal odd parameters $c_2$ and $f(\phi)$).
Our Onsager relation
\eqref{eq:explicitOnsager} is satisfied for the following special angles
\begin{equation}\label{eq:thetaRestrictions}
\theta = 0, \, \pi \qquad \forall c_1,\, c_2,\, f(\phi)\, .
\end{equation}
The case $\theta=0$ corresponds to the hard wall model (an IR Randall-Sundrum brane),
where the boundary theory
has a position-independent mass gap where $z=\infty$ is not part of our spacetime.
For the Hall conductivities we get $\sigma_{xy} = -2 f(\phi)$ and $\sigma_{yx} = 2 f(\phi)$.
In the other case, $\theta=\pi$, we get an UV Randall-Sundrum brane,
which we already discarded based on the discussion in section \ref{sec:example}.

Alternatively we can satisfy \eqref{eq:explicitOnsager} by choosing the following special
couplings\footnote{Additional solutions to the requirement that Onsager's relation holds:
\begin{equation}\label{eq:complexsolutions}
\qquad f(\phi) = -c_2 \pm i c_1  \qquad \forall c_1,\, c_2,\, \theta \, , \quad \quad \qquad
c_1 = \pm i(c_2 + f(\phi)) \qquad \forall c_2, \, f(\phi),\, \theta \, .
\end{equation}
These solutions give rise to complex couplings and can be immediately discarded.
In principle there may be also more complicated relations between the couplings
and $\theta$, which also lead to \eqref{eq:onsager} being satisfied.}

\begin{equation}\label{eq:fRestrictions}
f(\phi) = -c_2 \, .
\end{equation}
In this case we see from \eqref{ACT13} that only a (gravity) bulk
Maxwell-term and a Chern-Simons term on the AdS-boundary survive. There
is no contribution from the boundary $Q$ to the action, however $Q$ is present and
generates a position-dependent mass gap in the field theory. Here $Q$ lies in the
IR of the theory. This reconfirms our interpretation of the case of generic couplings being associated with spontaneous breaking of time-reversal invariance. In the standard holographic dictionary
terms in the action localized on the IR brane encode effects associated with the dynamics of symmetry breaking and confinement. They do not correspond to coupling constants in the field theory.
Adding a Chern-Simons term localized on the IR brane is adding dynamical breaking of time-reversal invariance to the holographic theory.
Therefore our Onsager relation should indeed hold whenever
there is no Chern-Simons term on $Q$.

\paragraph{Novel transport coefficients}
In addition to the conductivity, our system displayed a sequence of unusual transport coefficients
$\kappa_x,\, \kappa_y$ and
$\gamma_x,\,\gamma_y$. For example the $\kappa$'s correspond to a current driven by a magnetic field alone in the absence of any electric fields. In the hydrodynamic treatment of charge transport in time-reversal breaking theories in 2+1 dimensions such a term is impossible~\cite{Jensen:2011xb}. Deviations from the result
in~\cite{Jensen:2011xb} come from at least two distinctions. First, the treatment in~\cite{Jensen:2011xb} worked with an expansion, which considered electromagnetic fields to be of the same
order as first derivatives of hydrodynamic variables. In contrast
to that, in the present work we include electromagnetic background fields of zeroth order in derivatives. 
Second, we have at least one more quantity appearing in our low energy physics besides the standard hydrodynamic variables: the gradient of the condensate $\mathcal{O}$ that gives rise to our gap~\footnote{A similar effect has been studied in~\cite{Bhattacharyya:2008ji} in the context of the hydrodynamic description 
of an uncharged fluid.}. As the mass gap has a linear dependence
on the coordinate transverse to the defect, its gradient in the $y$-direction is a constant while all of its other derivatives vanish. With such a gradient it is trivial to write down an additional term in the constitutive relation for $J_x$:
\begin{equation} \label{eq:constitutiveJx}
J_x = \tilde{\kappa}_x F_{xy} (\partial_y \mathcal{O})
\end{equation}
which indeed gives us a constant $\kappa_x =\tilde{\kappa}_x  (\partial_y \mathcal{O})$ as long as the $y$ derivative of $\mathcal{O}$ is constant (assuming such a contribution is
consistent with the local version of the second law of thermodynamics~\cite{LL6}).
Note that since the gradient of $\mathcal{O}$ in the $x$ direction vanishes the corresponding term in $J_y$ gives no new contribution to the transport in perfect agreement with our result that $\kappa_y=0$. Similarly, $\gamma_y$ comes from a term in $J^t$ proportional to $F_{ty} (\partial_y \mathcal{O})$. Once more, the corresponding term $\gamma_x$ has to vanish as $\mathcal{O}$ has no gradient in the $x$ direction in perfect agreement with our results.

In contrast, the term $\kappa_t$ is already present in any theory with a mass gap and a non-vanishing Hall conductivity. Then $\kappa_t$ follows directly from a Chern-Simons term for the external electric fields which encapsulate the Hall conductivity. Note that in the standard quantum Hall effect with
a large magnetic field the definition of $\kappa_t=J^t/B$ coincides with the definition
for the filling fraction $\nu$ up to a factor of $(2\pi)$. However, the way in which we
set up our model, the relation $\nu/(2\pi) = \kappa_t$ is a result (which holds also
for $B\to 0$ and $J^t\to0$), rather than a definition.

In gapped systems only the external fields and the gap itself can contribute to
the low-energy excitations of the system. Thus for gapped quantum Hall systems
often an effective action of the form
\begin{equation}
S_\text{eff} = \int d^3x\, \hat\kappa  \, A\wedge F
  = \int d^3x \, \hat\kappa \, 2 \left (A_t B - A_x E_{y} + A_y E_{x}
  \right) + \dots \, ,
\end{equation}
with some coefficient $\kappa$ is considered. Now the currents $J^t$, $J^x$ and $J^y$
derived from this effective action have the same coefficient $\hat\kappa$.
From precisely these currents we extract $\kappa_t$ and $\sigma_{xy}$ in our model.
Therefore we expect $\kappa_t = \sigma_{xy}$, which is indeed what we find
in~\eqref{eq:JtCoefficients}.

Furthermore we also observe the equality $\gamma_y=\kappa_x$ in~\eqref{eq:JtCoefficients}.
This can be explained by a different term in the effective action accounting
for the condensate $\mathcal{O}$ which couples to the field strength $F$:
\begin{equation}
S_{\mathcal{O}} = \tilde \kappa \int d^3x \, \mathcal{O} dA \wedge * dA \,
 = \tilde\kappa \int d^3x \left( A_t E_y \partial_y \mathcal{O} +
  A_x B \partial_y \mathcal{O} \dots \right) \, .
\end{equation}
From this we see that $J^t/E_y$ and $J^x/B$ yield the same coefficient $\tilde \kappa$,
and thus $\gamma_y=\kappa_x$.

\section{$SL(2,\mathbb{Z})$ action on AdS/BCFT}
\label{sec:SL(2,Z)}

In this section, we consider the action of $SL(2,\mathbb{Z})$ on AdS/BCFT.
We are interested in this transformation because as a subgroup it has $\Gamma_0(2)$,
which is observed in $(2+1)$-dimensional electron systems
as the set of symmetries given in \eqref{DUA11}, \eqref{DUA12}, and \eqref{DUA13}.

\subsection{Review of $SL(2,\mathbb{Z})$ duality}
First, we ignore the boundary $Q$ and review $SL(2,\mathbb{R})$ symmetry of the $d=4$ $U(1)$ gauge theory.  Here, the $SL(2,\mathbb{R})$ symmetry should be broken to the discrete group $SL(2,\mathbb{Z})$ by quantum corrections. For superstring theory, this is the quantization of the charges of branes (see also~\cite{Bayntun:2010nx,Bayntun:2011bm}). It is found that this $SL(2,\mathbb{Z})$ symmetry is interpreted as the $SL(2,\mathbb{Z})$ duality including the mirror symmetry on the $d=3$ CFT side~\cite{Witten:2003ya}.

Defining the coupling constant $\tau =4\pi (c_2- i \, c_1)$,  the 4-dimensional action is defined as
\ba
I=\dfrac{1}{8\pi}\text{Im}\int \tau (F+i\tilde{F})\wedge *(F+i\tilde{F}) =\int\sqrt{-g}d^4x \Big(\dfrac{c_1}{2}F_{\mu\nu}F^{\mu\nu}-\dfrac{c_2}{2}F_{\mu\nu}\tilde{F}^{\mu\nu}\Big). \label{ACT41}
\ea
where $*$ is the 4-dimensional epsilon symbol in the general background and  $\tilde{F}_{\mu\nu}=\epsilon_{\mu\nu}{}^{\rho\lambda}F_{\rho\lambda}/2$.
 Remember that $**=-1$ for the operation on the 2-form. We chose the normalization of $\tau$ to reproduce the Abelian theory of~\cite{Hartnoll:2007ip} if we set  $\tau =\Theta/(2\pi)-2\pi/g^2i$.

We consider the $SL(2,\mathbb{R})$ transformation of $\tau$ while keeping the metric invariant as follows:
\ba
\tau ' =\dfrac{\hat a\tau +\hat b}{\hat c\tau +\hat d}, \label{COM12}
\ea
where $\hat a,\hat b,\hat c,\hat d\in \mathbb{R}$ and they satisfy $\hat a\hat d-\hat b\hat c=1$.

To consider the duality of $F$, we introduce the following quantity
\ba
\tilde{H}^{\mu\nu}\equiv -\dfrac{4\pi}{\sqrt{-g}}\dfrac{\delta I}{\delta F_{\mu\nu}}=4\pi (-c_1F^{\mu\nu}+c_2\tilde{F}^{\mu\nu}),
\ea
where $H$ can be interpreted as the Lagrange multiplier coupled with $F$.
This can also be written as
\ba
\mathcal{H}^{\mu\nu}=\bar{\tau}\mathcal{F}^{\mu\nu}, \label{EQU14}
\ea
where $\bar \tau$ is the complex conjugate of $\tau$ and
\ba
\mathcal{F}_{\mu\nu}=F_{\mu\nu}-i\tilde{F}_{\mu\nu},\quad \mathcal{H}_{\mu\nu}=H_{\mu\nu}-i\tilde{H}_{\mu\nu}.
\ea

Then, \eqref{EQU14} is invariant under the transformation \eqref{COM12} when simultaneously
transforming
\ba
\begin{pmatrix} \mathcal{H}_{\mu\nu}' \\ \mathcal{F}_{\mu\nu}' \end{pmatrix}=\begin{pmatrix}
\hat a& \hat b \\ \hat c& \hat d  \end{pmatrix} \begin{pmatrix}\mathcal{H}_{\mu\nu}  \\ \mathcal{F}_{\mu\nu} \end{pmatrix},
\ea
where the real part of this equation is given by
\ba
\begin{pmatrix} {H}_{\mu\nu}' \\ {F}_{\mu\nu}' \end{pmatrix}=\begin{pmatrix}
\hat a& \hat b \\ \hat c& \hat d  \end{pmatrix} \begin{pmatrix}{H}_{\mu\nu}  \\ {F}_{\mu\nu} \end{pmatrix}.
\ea

The EOM $\nabla_{\mu}\tilde{H}^{\mu\nu}=0$ and the Bianchi identity $\nabla_{\mu}\tilde{F}^{\mu\nu}=0$ are also invariant under the $SL(2,\mathbb{R})$ transformation.

A more interesting group is $SL(2,\mathbb{Z})$ in which we restrict the
parameters $\hat a,\hat b,\hat c,\hat d\in \mathbb{Z}$ satisfying $\hat a\hat d-\hat b\hat c=1$. This group can
be represented by two  matrices $S$ and $T$,
\ba
S=\begin{pmatrix} 0& 1 \\ -1 & 0 \end{pmatrix},\quad T=\begin{pmatrix} 1& 1 \\ 0 & 1 \end{pmatrix}.
\ea
Under $S,T$, the coupling constant transforms as
\ba
S:\ \tau' =-\dfrac{1}{\tau},\quad T:\ \tau' =\tau +1.
\ea
We can show that the above $S$ and $T$ action gives the proper transformation of the coupling constants after replacing $c_1$ and $c_2$ with $1/2g^2$ and $\Theta /(8\pi^2)$, respectively. The $S$ action maps $2\pi/g^2$ to $g^2/(2\pi)$ for $\Theta =0$ and The $T$ action acts as $\Theta\to\Theta +2\pi$.

\subsection{Transformation of conductivity under $S$ and $T$}

Now we introduce the boundary $Q$. Then we compute the transformation of the
conductivity under the action of $S$ and $T$.
We consider the case of $f(\phi)=0$ and the ansatz \eqref{ANS28} for convenience.

In terms of $\tilde{H}^{\mu\nu}$, the boundary condition \eqref{BOU18} at $Q$ becomes
\ba
n_{\mu}\tilde{H}^{\mu\nu}|_{Q}=0.
\ea
Under the $S$ action, this boundary condition is changed into
\ba
S:\ n_{\mu}\tilde{H}^{\prime \mu\nu}|_{Q}=n_{\mu}\tilde{F}^{\mu\nu}|_{Q}=0. \label{DIR311}
\ea
Note that the equation in the second equality describes the Dirichlet boundary condition at $Q$ of the field $F$ before the transformation instead of the Neumann boundary condition at $Q$.
The conductivity is given in terms of the dual fields $F',H'$. So $S$ turns the Neumann
into a Dirichlet boundary condition.
 Note that the boundary condition at $Q$ in terms of the dual fields $F',H'$ and $\tau'$ has
the same form as \eqref{BOU18}.

The current density derived from the GKPW relation is
\ba
&J^{t}=\dfrac{\sqrt{-g}\tilde{H}^{zt}}{2\pi}\Big|_{z=0}=\dfrac{H_{xy}}{2\pi}\Big|_{z=0},\quad J^{x}=\dfrac{\sqrt{-g}\tilde{H}^{zx}}{2\pi}\Big|_{z=0}=\dfrac{H_{yt}}{2\pi}\Big|_{z=0}, \nonumber \\
& J^{y}=\dfrac{\sqrt{-g}\tilde{H}^{zy}}{2\pi}\Big|_{z=0}=\dfrac{H_{tx}}{2\pi}\Big|_{z=0}.
\ea
After the $S$ transformation, the current and the Hall conductivity are written as
\ba
&J^{\prime \mu}=\dfrac{\sqrt{-g}\tilde{H}^{\prime z\mu}}{2\pi}\Big|_{z=0}=\dfrac{\sqrt{-g}\tilde{F}^{z\mu}}{2\pi}, \\
& \sigma'_{xy}=\dfrac{J^{\prime x}}{-F'_{t y}|_{z=0}}=\dfrac{F_{ty}|_{z=0}}{2\pi H_{yt}|_{z=0}},\quad \sigma'_{yx}=\dfrac{J^{\prime y}}{-F'_{t x}|_{z=0}}=\dfrac{F_{tx}|_{z=0}}{2\pi H_{tx}|_{z=0}}, \label{CON314} \\
& \kappa_t' =\dfrac{J^{\prime t}}{F'_{xy}|_{z=0}}=\dfrac{F_{xy}|_{z=0}}{-2\pi H_{xy}},
\ea
where we used the definition of the Hall conductivity. Note that the conductivity in \eqref{CON314} and $J^{\prime t}/B'$ become the same as \eqref{CUR112} if we replace $F,H,\tau$ with $F',H',\tau'$.

In the case of $\theta=\pi /2$, the Hall conductivity and $J^{\prime t}/B'$ are transformed as
\ba
&\sigma_{xy}'=2\Big(\dfrac{c_1^{\prime 2}}{c_2'}+c_2'\Big)=\dfrac{1}{2\pi}\dfrac{|\tau '|^2}{\mbox{Re}(\tau')} \nonumber \\
&=-\dfrac{1}{2\pi \mbox{Re}(\tau)}=-\dfrac{1}{8\pi^2 c_2}, \label{SDU416}
\ea
where we used the first equality of $\sigma_{xy}'$ in \eqref{CON314} and \eqref{CUR112}. We can also obtain the same formula using the second equality of $\sigma_{xy}'$. Solving the Dirichlet boundary condition $b=c=0$ in \eqref{DIR311}
\ba
\sigma'_{xy}=-\dfrac{E_y}{4\pi^2 (-2c_1c+2c_2E_y)}=-\dfrac{1}{8\pi^2 c_2}.
\ea

Then, we consider the $T$-transformation of the system.
Under the $T$ action, the boundary condition at $Q$ is changed into
\ba
T:\ n_{\mu}\tilde{H}^{\prime \mu\nu}|_{Q}=-4\pi c_1F^{\mu\nu}+(4\pi c_2+1)\tilde{F}^{\mu\nu}|_{Q}=0.
\ea
The condition above is consistent with the $T$-transformation of the coupling constant $c_2\to c_2+1/(4\pi)$. Remember that the gauge field $F_{\mu\nu}$ is not changed under the $T$-transformation.

For the case of $\theta=\pi/2$, the conductivity is transformed into
\ba
\sigma_{xy}'=2\Big(\dfrac{c_1^2}{c_2+1/(4\pi)}+c_2+\dfrac{1}{4\pi}\Big). \label{CON419}
\ea

We can repeat the previous analysis of the $SL(2,\mathbb{R})$ action, now for the generalized ansatz including the component $E_xx$ while imposing $f(\phi)=0$. It is easily shown that the conductivity and filling fraction are transformed by replacing $c_1,c_2,E_x,E_y,B$ with the dual parameters $c_1',c_2',E_x',E_y',B'$. The transformed conductivity is the same as \eqref{CON419} because of $f(\phi)=0$.

Similarly, we can perform the $SL(2,\mathbb{R})$ action for the case of the Dirichlet boundary condition at $Q$ \eqref{DIR31}. After replacing the parameters with the dual parameters, the Hall conductivity is given by
\ba
\sigma_{xy}'=-2c_1'\dfrac{B'}{E'}\cot\theta+2c_2'. \label{DIR420}
\ea

In summary, the new Hall conductivity \eqref{DIR420} is the same as the old
Hall conductivity \eqref{DIR34} with
the replacements of the parameters described above. We also notice that for $f(\phi)=0$, we can not impose the Onsager relation \eqref{eq:onsager} in the form $\sigma_{xy}'=\sigma_{yx}'$ after the transformation. This is because both the
$y$-component of the current and $\sigma_{yx}$ are also absent before the transformation as seen
from \eqref{eq:generalCurrent} (compare to \eqref{eq:explicitOnsager} at $f(\phi)=0$). Neither a current $y$-component nor $\sigma_{yx}'$ are generated after applying the transformation. Absence of $J^y$ is consistent with the current conservation on the BCFT side in presence of the boundary $y=0$ if there are no matter fields localized at $y=0$.

Upon quantization we consider the action of $SL(2,\mathbb Z)$ instead of $SL(2,\mathbb{Z})$.
It is interesting to compare the $SL(2,\mathbb{Z})$ transformation in this section with the
$\Gamma_0(2)$ transformation of the $d=(2+1)$ electron system mentioned in the introduction.
If $SL(2,\mathbb{Z})$ symmetry is broken to $\Gamma_0(2)$ via some mechanism, we can recover a \emph{duality transformation} similar to that of the $d=(2+1)$ electron system~\eqref{DUA11}-\eqref{DUA13}. The main difference between the two transformations being that $\Gamma_0(2)$ does not include the $S$ transformation.

The case of non-zero $f(\phi)$ is also interesting since such a system  can satisfy $\sigma_{xy}=\sigma_{yx}$. However, the $SL(2,\mathbb{Z})$ transformation in the (gravity) bulk acts on the gauge fields at $Q$ and the action for $f(\phi)$ should be specified explicitly. In this case one also has to investigate the gauge invariance of the Chern-Simons term at $Q$ under the $SL(2,\mathbb{Z})$ action.


\section{Stringy realization}
\label{sec:stringy}
According to \cite{Fujita:2011fp}, the type IIA string theory on $AdS_4\times \mathbb{CP}^3$ can realize the AdS/BCFT using the orientifold 8-planes.
In string frame, the type IIA background~\cite{Aharony:2008ug} is
\ba
&g=R^2\Big(r^2dx^{\mu}dx_{\mu}+\dfrac{dr^2}{r^2}+4ds^2_{\mathbb{CP}^3}\Big), \nonumber \\
&e^{2\phi}=\dfrac{4R^2}{k^2},\quad F_4=-\dfrac{3}{2}kR^2\text{vol}(AdS_4),\quad F_2=\dfrac{k}{2R^2}\omega,
\ea
where we set $\alpha'=1$, $\omega$ is the K{\"a}hler form on $\mathbb{CP}^3$, and $R^2=2^{1/2}\pi\sqrt{N/k}$.
Here, $ds^2_{\mathbb{CP}^3}$ and the potential $A$ for $F_2$ are given by
\ba
&ds^2_{\mathbb{CP}^3}=d\xi^2+\cos^2\xi\sin^2\xi \Big(d\psi+\frac{\cos\alpha_1}{2}d\phi_1-\frac{\cos\alpha_2}{2}d\phi_2\Big)^2 \nonumber \\
&+\frac{1}{4}\cos^2\xi(d\alpha_1^2+\cos^2\alpha_1d\phi_1^2)+\frac{1}{4}\sin^2\xi\Big(d\alpha_2^2+\cos^2\alpha_2 d\phi_2^2\Big),  \\
&2A=k(\cos 2\xi d\psi +\cos^2\xi \cos \alpha_1 d\phi_1 +\sin^2\xi \cos\alpha_2 d\phi_2),
\ea
where $0\le \xi \le \pi/2$, $0\le \alpha_i\le \pi$, $0\le \phi_i\le 2\pi$, and $0\le \psi\le 2\pi$.
 We then introduce the two orientifold 8-planes placed at $y=0$ and $y=L$ which are extending to the directions except for $y$. For $O8^--O8^+$ system as an example, a half of the original supersymmetry is preserved as proven in~\cite{Fujita:2009kw}. Moreover, both NSNS and RR tadpole are canceled. For the $O8^--O8^-$ system, on the other hand, 16 D8-branes are needed to cancel the RR charge source. The theory has vanishing tension $T=0$ since the NS tadpole is also canceled. Correspondingly, this setup can only realize the special $\theta=\pi/2$ case.

First, we review the derivation of the FQHE for the case of no O8 planes~\cite{Hikida:2009tp}. We show that we can construct the FQHE in terms of $RR$-flux in the bulk $AdS_4$.
The following ansatz is assumed:
\begin{align}
 F_2=\f{k}{2R^2}\omega + F' ~,\qquad
 \ti{F}_4=-\f{3kR^2}{2}\ep_{AdS_4}+\f{\pi }{R^2}F_{ext}\we \omega ~, \label{ANS54}
\end{align}
and the 3-form $H_3$ is assumed to have the indices only in the $AdS_4$ directions. Moreover, $F'$ and $F_{ext}$ have indices in the $AdS_4$ direction. According to~\cite{Hikida:2009tp}, a combination of these fluxes becomes a massless gauge field. In this ansatz,
\begin{align}
 *F_2=\f{kR^2}{4}\ep_{AdS_4}\we\omega^2+*_4F'\we \f{\omega^3}{6} ~,\qquad
 *\ti{F}_4=-\f{k}{4R^2}\omega^3 +\f{\pi }{2R^2}*_4F_{ext}\we \omega^2 ~,
\end{align}
where $*$ and $*_4$ is the Hodge duals in the total  10-dimensional spacetime and $AdS_4$ spacetime, respectively.

The EOM of fluxes is written at the linearized level
as follows:
\begin{align}
 dF'=0 ~,\qquad dH_{3}=0 ~, \qquad
 dF_{ext}=-\f{k}{2\pi}H_3 ~, \qquad
 d*_4F'=\f{3k}{2R^2}H_3 ~, \no
 d*_4F_{ext}=0 ~,\qquad
 \f{1}{g_s^2}d*_4H_3=-\f{3\pi k}{2R^4}*_4F_{ext}-\f{3k}{2R^2}F' ~. \label{eomr}
\end{align}
Then, the following mode becomes a massless 2-form field strength:
\be
F'=-\f{\pi }{R^2}*_4F_{ext} ~,\ \ \ H_3=0. \label{masslessm}
\ee
The EOM (\ref{eomr}) can be solved beyond the linear order approximation.

The background is assumed to include 2-form flux $\int_{\mathbb{CP}^1} B_2/(2\pi)^2=M/k$. If the massless mode is considered, it is found that the type IIA
action is reduced to
 \be S_{ext}=-\f{R^2}{12\pi^2 }\int_{AdS_4}
F_{ext}\we
*F_{ext}-\f{M}{4\pi k}\int_{AdS_4} F_{ext}\we F_{ext}
~,\label{gauge} \ee where
the second term is derived from the C-S term of the type IIA supergravity
$-\f{1}{4\kappa^2}\int
B_2\we dC_3\we dC_3.$ In addition, the second term can be integrated by parts. And it gives a boundary C-S term $\int A_{ext}\we F_{ext}$ in the AdS/CFT correspondence.

Here, $A_{ext}$ is the external gauge field which appeared in the FQHE. So, the fractionally quantized Hall conductivity is obtained as the coefficient of the external field in the GKPW relation:
\ba
J^x=\dfrac{\delta S_{ext}}{\delta A_{ext}}=\dfrac{M}{2\pi k}*F_{ext}.
\ea
Thus, the fractionally quantized Hall conductivity becomes
\be \sigma_{xy}=-\dfrac{M}{2\pi k}=-\dfrac{M}{k}\cdot\dfrac{e^2}{h}, \ee
where the electron charge $e$ and Planck constant $\hbar$ are restored.
Thus, the ABJM theory with $M$ 2-form flux can be the model of the fractional quantum Hall effect. As pointed out in~\cite{Hikida:2009tp},
 the case of no fractional branes $M=0$ is also interesting. Since the electric-magnetic
 duality transformation acting on the Maxwell term in the (gravity) bulk inverts the coupling constant of the Maxwell term $g\sim (k/N)^{1/4}$, this transformation exchanges the rank $N$ and the level $k$ of the ABJM theory.

Second, we place two O8-planes at $y=0$ and $y=L$. We choose the case of two $O8^-$-planes and 16 D8-branes or the $O8^{+}-O8^-$ system. As known in the type IIA string theory on orientifolds~\cite{Dabholkar:1996pc}, we can derive the action of the orientifolds on the fields of the supergravity. Remember that NSNS 2-form $B_{2}$ is odd under the orientifold and the other NS fields are trivial under it. According to~\cite{Hanany:2000fq}, we have the following action of the O8-plane on the RR forms:
\ba
C_1\to C_1,\quad C_3\to -C_3, \label{RR511}
\ea
where the above sign is in addition to a sign coming from the tensor transformation property of the RR forms. Note that the orientifold action
 preserves the C-S term of the type IIA theory.

Returning to the ansatz \eqref{ANS54}, we notice that $F'$ is even and $F_{ext}$ is odd under the orientifold since $\omega$ is even under the orientifold action. It is found that the condition \eqref{RR511} and the odd property of $B_{NS}$ do not change the massless solution \eqref{masslessm}. Thus, we derive the same 4-dimensional action \eqref{gauge}.

To compute the Hall conductivity using AdS/BCFT, we should replace the coefficient $c_1,c_2$ in \eqref{ACT13} with
\ba c_1=-\dfrac{R^2}{12\pi^2},\quad c_2=-\dfrac{M}{4\pi k},\quad f(\phi)=0. \ea
Here, the angle of the boundary $Q$ is $\theta=\pi/2$.
Note that since $F_{ext}$ is odd under the orientifold, we should choose the Dirichlet boundary condition at $Q$ \eqref{DIR31}.
Remember that the integration of the topological term gives Chern-Simons term of different signs at each boundary. Thus, we can show that the boundary conditions at both $y=0$ and $y=L$ become the same and are given by
\ba
c=b=0, \quad F_{tx}=0.
\ea

Substituting the above formula into \eqref{DIR34}, we obtain the Hall current with filling fraction $\nu=M/k$ as follows:
\ba
J^t=-\dfrac{M}{2\pi k}B,\quad J^x=-\dfrac{M}{2\pi k}E_y,
\ea
After the $S$-transformation and $T$-transformation and using \eqref{DIR420}, the Hall conductivity becomes
\ba\label{eq:STsigma0}
S:\ \sigma_{xy}=\dfrac{9Mk}{2\pi(2Nk+9M^2)},\quad T:\sigma_{xy}=-\dfrac{M}{2\pi k}+\dfrac{1}{2\pi},
\ea
and
\ba\label{eq:STsigma}
S T^2 S: \sigma_{xy}=-\dfrac{1}{2\pi}\dfrac{18M^2+9Mk+4Nk}{36M^2+36Mk+9k^2+8Nk}.
\ea
The inverse of above transformation $ST^{-2}S$ corresponds to the third of the three \emph{duality transformations}, \eqref{DUA13}.
Note that $T: \sigma_{xy}\to\sigma_{xy} + 1/(2\pi)$ in \eqref{eq:STsigma},
which is the first of the three \emph{duality transformations}, \eqref{DUA11}.
The second one \eqref{DUA12}, $\sigma_{xy}\to 1/(2\pi)-\sigma_{xy}$, corresponds to the $T^{-1}$-transformation and subsequent scaling $\sigma_{xy}\to -\sigma_{xy}$.
However, the non-trivial action of $S$ given by \eqref{eq:STsigma0} is new in this context.

\section{Discussion}
\label{sec:discussion}
In this paper, we analyzed the response of a conserved current to external electromagnetic fields in the AdS/BCFT correspondence recently proposed in \cite{Takayanagi:2011zk}. Starting
out bottom-up, we proposed an Abelian gauge theory in the 4-dimensional gravity theory with a boundary $Q$ at different angles $\theta$, see figure~\ref{fig:adsbcft}. At this boundary $Q$ we considered the Neumann and the Dirichlet boundary condition, and derived the Hall conductivity of the
holographically dual $(2+1)$-dimensional theory.

We note that our Hall conductivities are nonzero even in the absence of any large
external magnetic field, see e.g.~\eqref{eq:sigmaxyThetaPi/2}.
Therefore this setup may be interpreted to model the
anomalous Hall effect (or intrinsic Hall effect)~\cite{nsomo} alongside the standard
quantum Hall effect.
Our Hall conductivity transforms in a particular way under the group $SL(2,\mathbb{Z})$. This group is related to the well-known \emph{duality transformations} \eqref{DUA11}--\eqref{DUA13}
of the conductivity in a $(2+1)$-dimensional electron gas. As mentioned in the introduction, these transformations are generated by the group $\Gamma_0(2)$ which is a subgroup of $SL(2,\mathbb{Z})$. In section \ref{sec:SL(2,Z)} we derived the transformation law~\eqref{DIR420} for the conductivity under the larger group $SL(2,Z)$ via holography. This may be interpreted as a \emph{generalized duality transformation} of the $(2+1)$-dimensional BCFT (which is a generalization of the known duality transformation generated by the smaller group $\Gamma_0(2)$).
For the example $\theta=\pi/2$, we found that the conductivity on the BCFT side has a non-trivial transformation under $SL(2,\mathbb{Z})$ since the Hall conductivity is of the form $|\tau |^2/(2\pi\mbox{Re}(\tau))$ as seen from \eqref{SDU416}.

We also considered the $d=10$ type IIA string embedding of our system.
After dimensional reduction to $d=4$, we obtain the Abelian action of the massless fields. It is found that the gauge field satisfies the Dirichlet boundary condition at the boundary $Q$ which we analyzed in section 3. It was found that the $d=4$ Abelian theory realizes the FQHE
with vanishing longitudinal conductivity. As we anticipated in the introduction, 
the $SL(2,\mathbb{Z})$ transformation for this string model indeed acts on $\sigma_{xy}$ 
differently from the transformation \eqref{DUA14}. Instead we find the 
transformation behavior described in~\eqref{DIR420}.

Alongside the Hall conductivity we also find several potentially novel transport
coefficients $\kappa_x = J^x/B,\, \kappa_y= J^y/B = 0$
and $\gamma_x = J^t/E_x = 0,\, \gamma_y = J^t/E_y =\kappa_x$. The relations
among these coefficients and their interpretation are discussed in section~\ref{sec:conductivities}.
Imposing Onsager relations puts further restrictions on the coupling constants and the
angle $\theta$ in our model as discussed in section~\ref{sec:conductivities} as well.
Another coefficient $\kappa_t = J^t/B = \sigma_{xy}$ is found, which turns out to be related
to the filling fraction $\nu = 2\pi \kappa_t$. Recall that we do not take
$J^t/B$ as a definition of the filling fraction $\nu$, because our Hall conductivity
in general is nonzero at vanishing magnetic field $B=0$.

Our model is capable of producing many different filling fractions $\nu$,
see e.g. \eqref{eq:sigmaxyThetaPi/2}, \eqref{eq:sigmaxyTheta0}. However, we
highlight the case $\nu=5/2$ which is of particular interest to the condensed matter
community. Systems with $\nu=5/2$ are thought to give rise to non-Abelian
quasi-particles which may be utilized to engineer braided states for quantum
computers~\cite{kamnayak-2008-80}. Our model realizes
$\nu=5/2$ at vanishing Chern coupling $k=4\pi f(\phi)=0$ on the boundary $Q$,
with the other couplings being $l=4\pi c_2 = 2$ with the boundary $Q$ at
an angle $\theta=\pi/2$, see figure \ref{fig:adsbcft}.

Looking out into the future, it would be interesting to study the case
of two boundaries $Q_1$ and $Q_2$ at
possibly distinct angles $\theta_1$ and $\theta_2$, respectively. Furthermore, we
have left a few ends loose after the present work: One should explore
the action of $SL(2,\mathbb{Z})$ on the more general model with
non-vanishing $f(\phi)$. Also a non-trivial time-dependence
for the gauge field ansatz analogous to~\cite{Karch:2010kt} may give deeper insights
into the nature of the boundary CFT which we have been studying here.
Finally, our stringy model was limited to realize the boundary configuration
with $Q$ at an angle $\theta=\pi/2$, but there may be a whole family
of stringy models which realize other values of $\theta$.

\bigskip
\noindent {\bf Acknowledgments:}  We would like to thank Y. Nakayama, T. Nishioka, T. Takayanagi, D. T. Son, E. Tonni, and Herman Verlinde for discussions and comments. MF is supported by a Japan Society for the Promotion of Science (JSPS) Postdoctoral Fellowship program for Research Abroad. MK and AK are supported in part by the U.S. Department of Energy under Grant No.~DE-FG02-96ER40956.

\end{document}